\documentclass[epj,10p]{svjour}
\usepackage{epsfig}
\newcommand{\be}{\begin{equation}}
\newcommand{\ee}{\end{equation}}
\newcommand{\ba}{\begin{eqnarray}}
\newcommand{\ea}{\end{eqnarray}}
\usepackage{graphicx}
\usepackage{dcolumn}
\usepackage{bm}
\usepackage{hyperref}
\usepackage{amssymb}
\usepackage{hyperref}
\usepackage{graphicx}
\usepackage{dcolumn}
\usepackage{bm}
\usepackage{hyperref}
\RequirePackage[normalem]{ulem} 
\RequirePackage{color}\definecolor{RED}{rgb}{1,0,0}\definecolor{BLUE}{rgb}{0,0,1} 

\begin{document}
\title{Ground-state correlation properties of charged bosons trapped in strongly anisotropic harmonic potentials}

\author{Przemys\l aw Ko\'scik\inst{1}, Anna Okopi\'nska\inst{1}}

\institute{Institute of Physics,   Jan Kochanowski University\\
ul. \'Swi\c{e}tokrzyska 15, 25-406 Kielce, Poland}

\abstract{We study systems of a few charged bosons contained within
a strongly anisotropic harmonic trap. A detailed examination of the
ground-state correlation properties of two-, three-, and
four-particle systems is carried out within the framework of the
single-mode approximation of the transverse components. The linear
correlation entropy of the quasi-1D systems is discussed in
dependence on the confinement anisotropy and compared with a strictly 1D limit. Only at weak
interaction the correlation properties depend strongly on the anisotropy parameter.
}

\authorrunning{:}
 \titlerunning{}

\maketitle
\section{Introduction}
Systems of few interacting particles confined by an external
potential are of increasing interest in view of their application to
model various nanostructures fabricated in an artificial way. The
Schr\"{o}dinger equation for harmonically trapped particles which
interact through a Coulomb potential can be used to simulate
different physical systems, such as semiconductor quantum
dots~\cite{fab} or electromagnetically trapped ions \cite{ion}.
Besides of fermionic systems, also the bosonic ones have been thus
considered in various theoretical contexts both in the 3D
\cite{blu,bao,nboson} and 2D \cite{gonz} case. The  properties of
quasi-1D systems are, however, rather rarely studied, except for the
models of bosons with a contact potential
\cite{int6,tg0,1,2,3,4,1d}. In this work, we discuss the case of $N$
identically charged spinless bosons confined in a quasi-1D trap.
Such a trapping potential may be experimentally realized using
highly anisotropic harmonic traps where the radial confinement is
much tighter than the axial one.

We discuss the effects of both the number of particles
and the control parameters of the system on the correlation
properties.  We consider various characteristics such as
single-particle reduced density matrix, single-particle density, and
linear entropy. In particular, the
influence of the anisotropy on the correlations within a quasi-1D
structure is investigated.

Our paper is organized as follows. In Section \ref{1} we present the model and provide 
an analytical formula for the effective interaction potential. Section \ref{2} surveys the quantities we use to
characterize correlations in the system. In section \ref{1D} the limit of
$\epsilon\rightarrow \infty$ is discussed. The results are presented in Section
\ref{results} and a summary of our conclusions is given in Section
\ref{4}.

\section{Effective Hamiltonian}\label{1}

\begin{figure}[h]
\begin{center}
\includegraphics[width=0.49\textwidth]{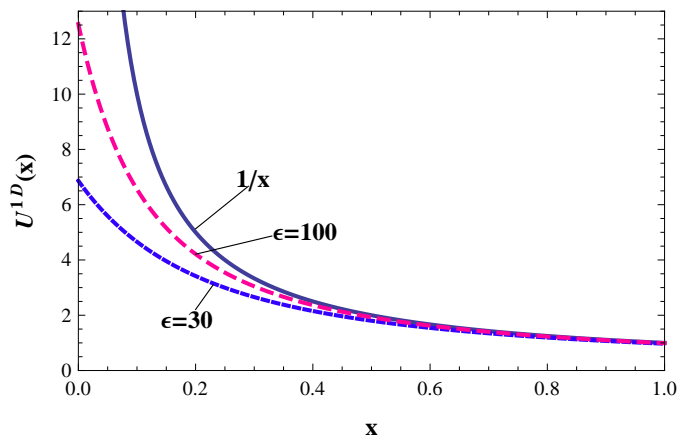}
\end{center}
\caption{\label{fffklolofghklg:beh}Effective interaction potential
(\ref{eff}) as a function of  $x=x_{2}-x_{1}$ for the anisotropy
parameter $\epsilon=30,100$ compared with the Coulomb potential
$x^{-1}$.}\end{figure}

Consider a system of $N$ Coulombically interacting particles trapped in a 3D axially-symmetric harmonic potential with a Hamiltonian given by
\begin{eqnarray} H=\sum_{i=1}^{N}[-{\hbar^2 \bigtriangledown_{i}^2\over 2 m}
+ {m\over 2}(\omega_{x}^{2} x_{i}^2+\omega_{\perp}^2\rho _{i}^2)]
 +\sum_{i<j}{\gamma\over |\textbf{r}_{i}-\textbf{r}_{j}|},\label{ham}\end{eqnarray}
where $\rho_{i}=\sqrt{y_{i}^2+z_{i}^2}$. After the scaling
$\textbf{r}\mapsto \sqrt{\hbar\over {{m\omega_{x}}}}\textbf{r}$,
$E\mapsto {\hbar \omega_{x}E}$, the Schr\"{o}dinger equation takes
the form \be
H\Psi(\textbf{r}_{1},\textbf{r}_{2},...,\textbf{r}_{N})=E\Psi(\textbf{r}_{1},\textbf{r}_{2},...,\textbf{r}_{N}),\label{ghj}
\ee with
\begin{eqnarray} H= \sum_{i=1}^{N}[-{\bigtriangledown_{i}^2\over
2}+{1\over 2} x_{i}^2+{1\over 2}\epsilon^2 \rho_{i}^2]
 +\sum_{i<j}{g\over
|\textbf{r}_{i}-\textbf{r}_{j}|} \label{hamrt}.\end{eqnarray} The dimensionless coupling $g={\gamma}\sqrt{{m\over
\omega_{x}\hbar^{3}}}$ represents the ratio of the Coulomb interaction to the longitudinal trapping  energy
scale and the dimensionless parameter
$\epsilon={\omega_{\perp}\over \omega_{x}}$ measures the anisotropy
of the trap. We focus our attention on the
strong anisotropy case, $\epsilon\gg 1$, when the system becomes quasi-1D. 
In this case the particles may be assumed to stay in the
lowest energy state of the tranverse Hamiltonian
$H_{\perp}=-{\bigtriangledown_{\rho}^2\over 2}+  {1\over
2}\epsilon^2 \rho^2$ and the one-mode approximation is justified. The $N-$body wave function may be taken in the form
\begin{eqnarray}
\Psi(\textbf{r}_{1},\textbf{r}_{2},...,\textbf{r}_{N})\cong\psi(x_{1},x_{2},...,x_{N})\Pi_{i=1}^{N}\varphi(y_{i})\varphi(z_{i}),\label{ddd}\end{eqnarray}
where $ \varphi(z)=({\epsilon\over \pi})^{{1\over
4}}e^{-\frac{\epsilon z^2}{2}}$ and $\psi$ is assumed to be a real
function. After substituting (\ref{ddd}) into (\ref{ghj}),
multiplying it from the left by $
\varphi(y_{1})\varphi(z_{1})...\varphi(y_{N})\varphi(z_{N})$ and
integrating over $y_{1},y_{2},...,y_{N}$, $z_{1},z_{2},...,z_{N}$ we
arrive at \be
H_{1D}\psi(x_{1},x_{2},...,x_{N})=E_{1D}\psi(x_{1},x_{2},...,x_{N}),\label{EOM1}\ee
where the quasi-1D Hamiltonian has a form
\begin{eqnarray} H_{1D}= \sum_{i=1}^{N}[-{1\over 2}{\partial^2\over \partial x_{i}^2}+{1\over 2} x_{i}^2]
 +\sum_{i<j} g U_{1D}(x_{i},x_{j})+N\epsilon.\label{quasio}\end{eqnarray}
The effective interaction potential
\begin{eqnarray}U_{1D}(x_{1},x_{2})=\int
{[\varphi(y_{1})\varphi(y_{2})\varphi(z_{1})\varphi(z_{2})]^2\over
|\textbf{r}_{1}-\textbf{r}_{2}|}dy_{1}dy_{2}dz_{1}dz_{2},\end{eqnarray}
is calculated to be given by \be
U_{1D}(x_{1},x_{2})=\sqrt{{\epsilon\pi\over 2}}e^{{\epsilon
(x_{2}-x_{1})^2\over 2}}(1-erf[{\sqrt{\epsilon\over
2}}|x_{2}-x_{1}|]),\label{eff}\ee where $erf(z)$ is the error
function. In Fig.\ \ref{fffklolofghklg:beh} the effective potential
(\ref{eff}) is compared with the pure Coulomb potential for
different values of $\epsilon$. We can notice that the larger is the
value of $\epsilon$, the closer to the origin does the effective
potential begin to exhibit Coulomb behaviour.

\begin{figure}[h]
\begin{center}
\includegraphics[width=0.49\textwidth]{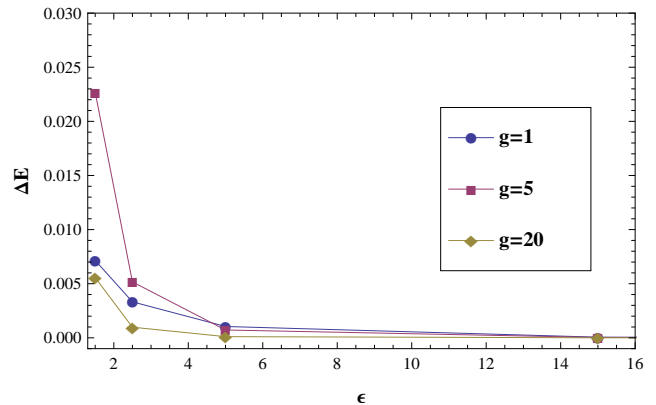}
\end{center}
\caption{\label{fffklolofghk3lg:beh} Relative ground-state energy
error \mbox{$\Delta E=|E-E_{1D}|/E$} as a function of $\epsilon$ for
$N=2$ and $g=1,5,20$.}\end{figure}

We have tested the applicability of the single mode approximation
for the two-particle system ($N=2$) by comparing the ground-state
energy $E_{1D}$ obtained from the 1D Hamiltonian (\ref{quasio}) with
the energy $E$ determined from the full 3D Hamiltonian
(\ref{hamrt}). The relative error  $\Delta E=(|E-E_{1D}|)/E_{0}$ as
a function of $\epsilon$ is shown in Fig. \ref{fffklolofghk3lg:beh}
for different values of the coupling constant.
One can conclude that anisotropy ratios $\epsilon\gtrsim 5$ are sufficiently large for employing the single mode approximation.

\section{Correlation characteristics}\label{2}
A basic tool to investigate two-body correlations in the system is
the one-particle reduced density matrix (RDM)\cite{redu} for spinless particles defined as
\be \rho(\textbf{r},\textbf{r}^{'})=\int\ldots\int \Psi(\textbf{r},\textbf{r}_{2},...,\textbf{r}_{N})\Psi(\textbf{r}^{'},\textbf{r}_{2},...,\textbf{r}_{N})d^3\textbf{r}_{2}...d^3\textbf{r}_{N}.\ee
In the one-mode approximation (\ref{ddd}) the RDM
takes the form  \be
\rho(\textbf{r},\textbf{r}^{'})=\varphi(y)\varphi(y^{'})\varphi(z)\varphi(z^{'})\rho_{1D}(x,x^{'}),\ee
with the 1D effective RDM given by \be \rho_{1D}(x,x^{'})=\int\ldots\int
\psi(x,x_{2},...,x_{N})\psi(x^{'},x_{2},...,x_{N})dx_{2}...dx_{N}
\label{RDM}.\ee  
The effective RDM can be represented in the Schmidt form \be
\rho_{1D}(x,x^{'})=\sum_{l=0}^{\infty}\lambda_{l}v_{l}(x)v_{l}(x^{'}),\label{ex}\ee
where $\{v_{l}(x)\}$ are the natural orbitals and their occupancies
$\{\lambda_{l}\}$. We will concentrate on discussing the linear
entropy
 \be L=1-\int\int
\rho_{1D}(x,x')^2dx' dx\label{lnearll}.\ee which can be expressed in
terms of $\lambda_{l}$ as $L=1-\sum_{l}\lambda_{l}^2$. It
gives indication of the spread of terms in the Schmidt decomposition
(\ref{ex}) and is one of the popular measures of correlation
~\cite{lin3,helium,manz,kosPLA}.

\section{Strictly 1D limit}\label{1D}
In the strictly 1D limit of $\epsilon\rightarrow\infty$, the
interaction potential of the considered system, $g/|x_{i}-x_{j}|$,
diverges at $x_{i}=x_{j}$ for any finite $g$. This causes
divergences at short distances in calculating the energy of bosonic
systems. Usually those divergences are cured by performing
calculation for an anisotropic 3D system with finite $\epsilon$ and
observing the limiting behavior at $\epsilon\rightarrow \infty$.
However, as noticed recently \cite{astr}, the calculation may be
performed directly for the 1D system since its ground-state
wavefunction can be related via Bose-Fermi mapping to the lowest
energy antisymmetric $N$-particle wavefunction $\psi_{F}$ as \be
\psi(x_{1},x_{2},..,x_{N})=|\psi_{F}(x_{1},x_{2},..,x_{N})|.\label{lld}\ee
Therefore, the system of bosons gets fermionized and the ultraviolet
divergencies are cured in a natural way when determining $\psi_{F}$
by the standard configuration interaction method.

\section{Results and discussions}\label{results}
We shall consider the ground-state of many-body systems with the
number of particles $N=2,3$ and 4. In order to reveal qualitatively
the nature of correlations in the quasi-1D  limit, we first discuss
the case of large anisotropy, $\epsilon=30$. The ground-state
$N$-particle wave function of the Hamiltonian (\ref{quasio}) is calculated with the quantum
diffusion algorithm \cite{difu} and used to determine $\rho(x,x^{'})$  by numerical integration of Eq. (\ref{RDM}).
\begin{figure}[h]
\begin{center}

\begin{picture}(6,6)

\put(72.5,1) {\small $N=4$}
\put(-5.,1) {\small $N=3$}

\put(-80.0,1) {\small  $N=2$}
\put(-125.0,-23) {\scriptsize $g=1$}
\put(-125.0,-100) {\scriptsize $g=2$}
\put(-125.0,-174) {\scriptsize $g=5$}

\end{picture}

\includegraphics[width=0.143\textwidth]{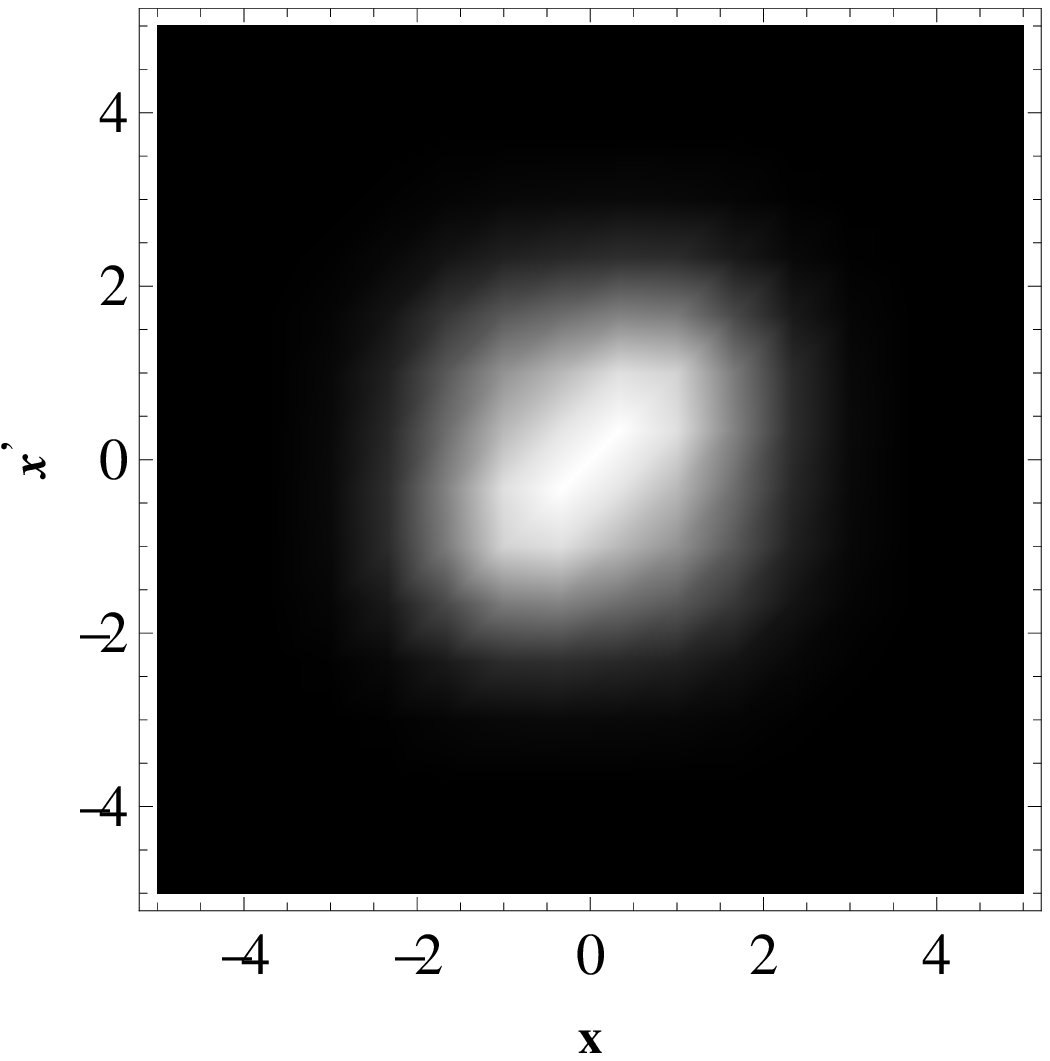}
\includegraphics[width=0.143\textwidth]{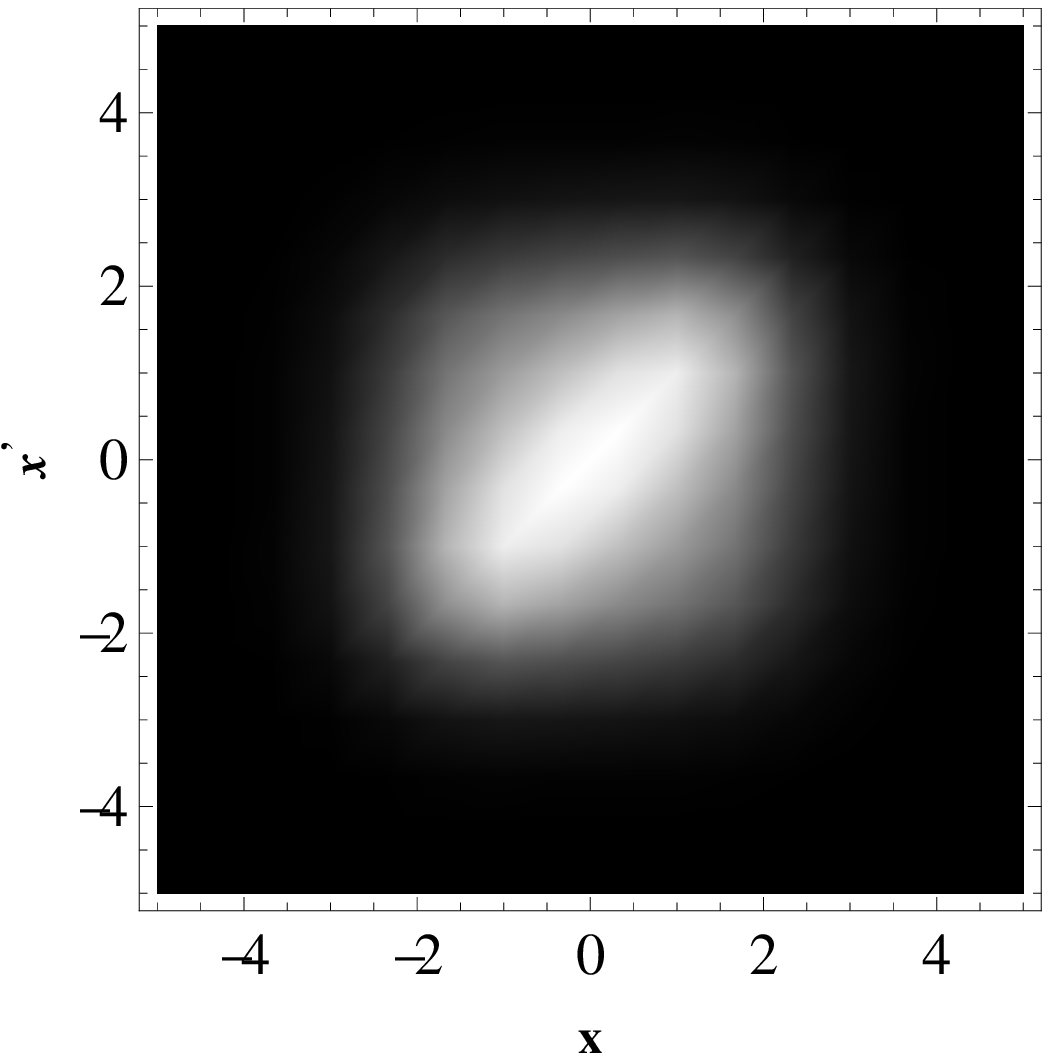}
\includegraphics[width=0.143\textwidth]{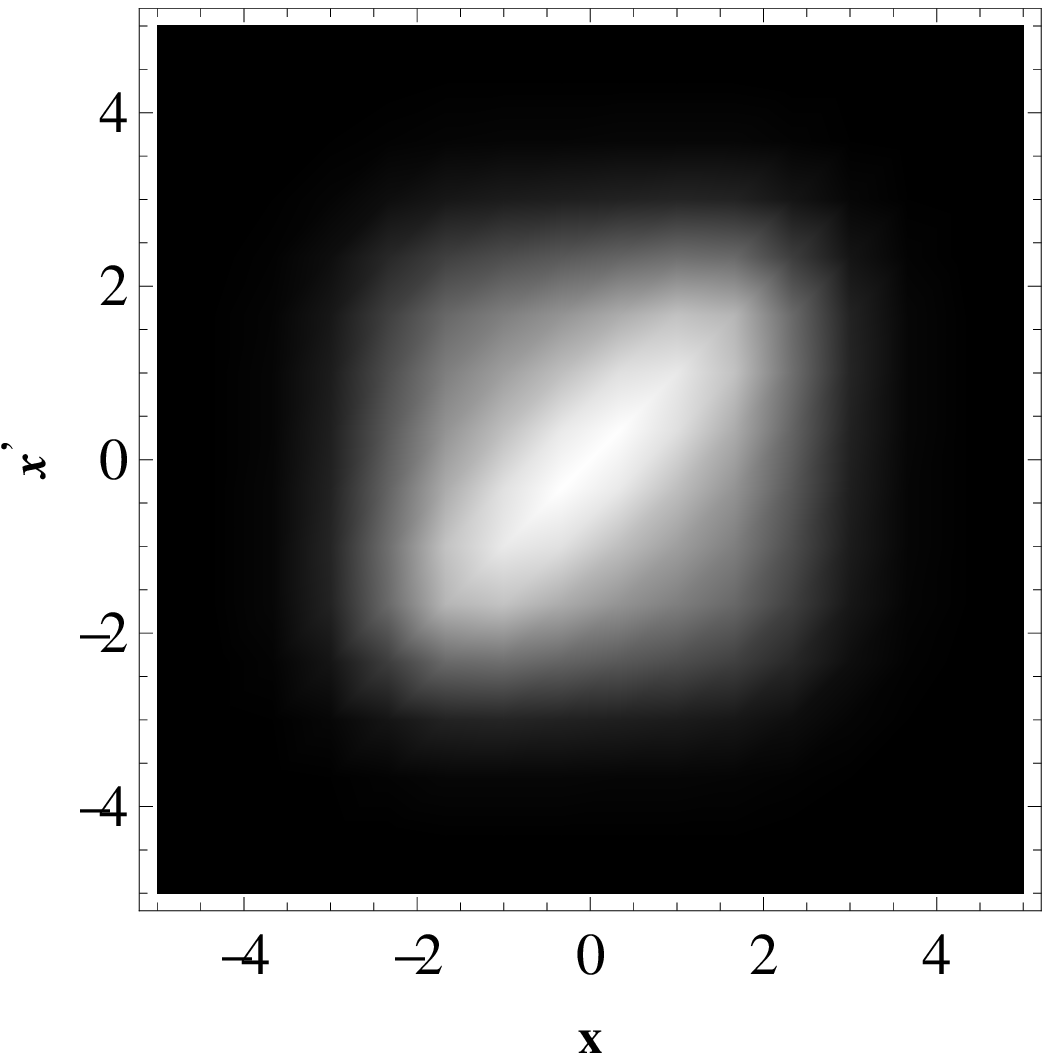}
\includegraphics[width=0.143\textwidth]{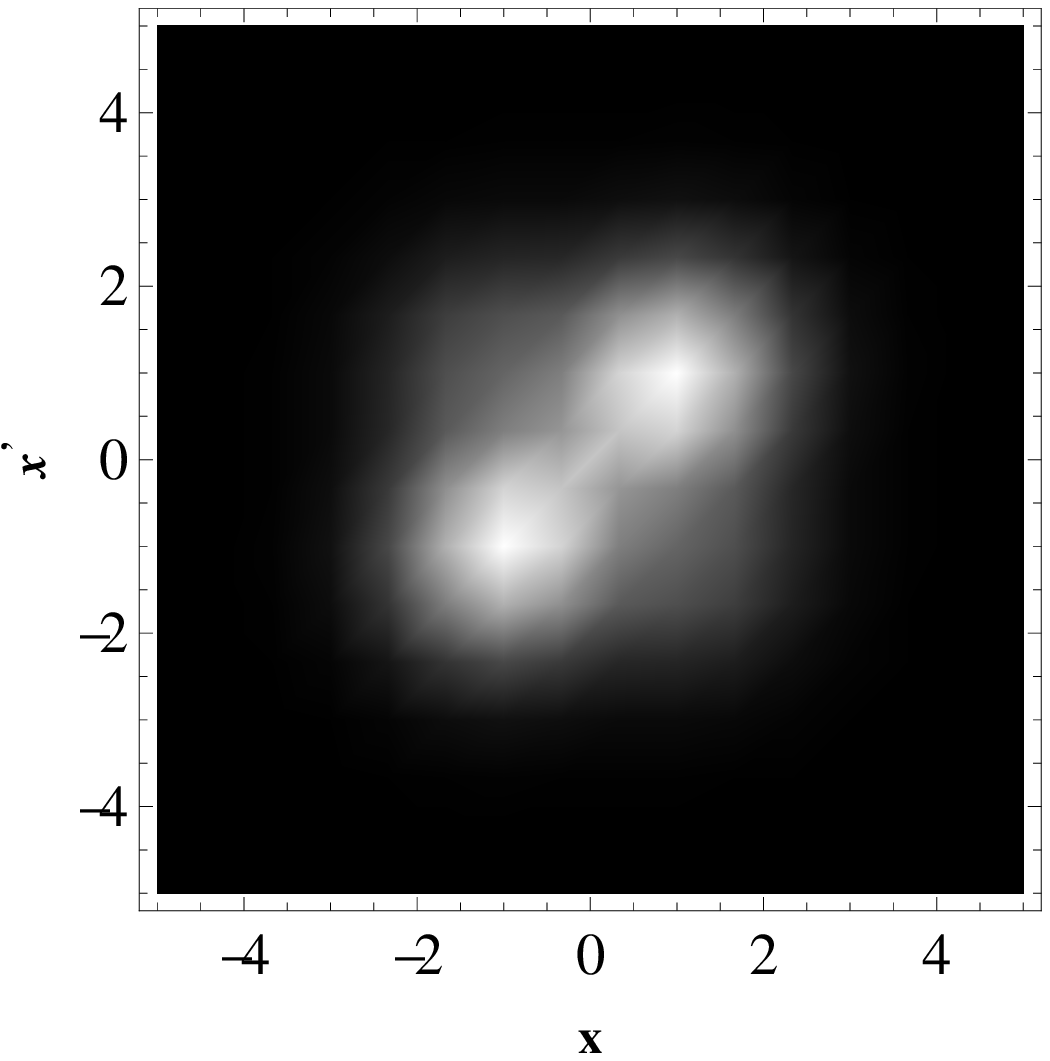}
\includegraphics[width=0.143\textwidth]{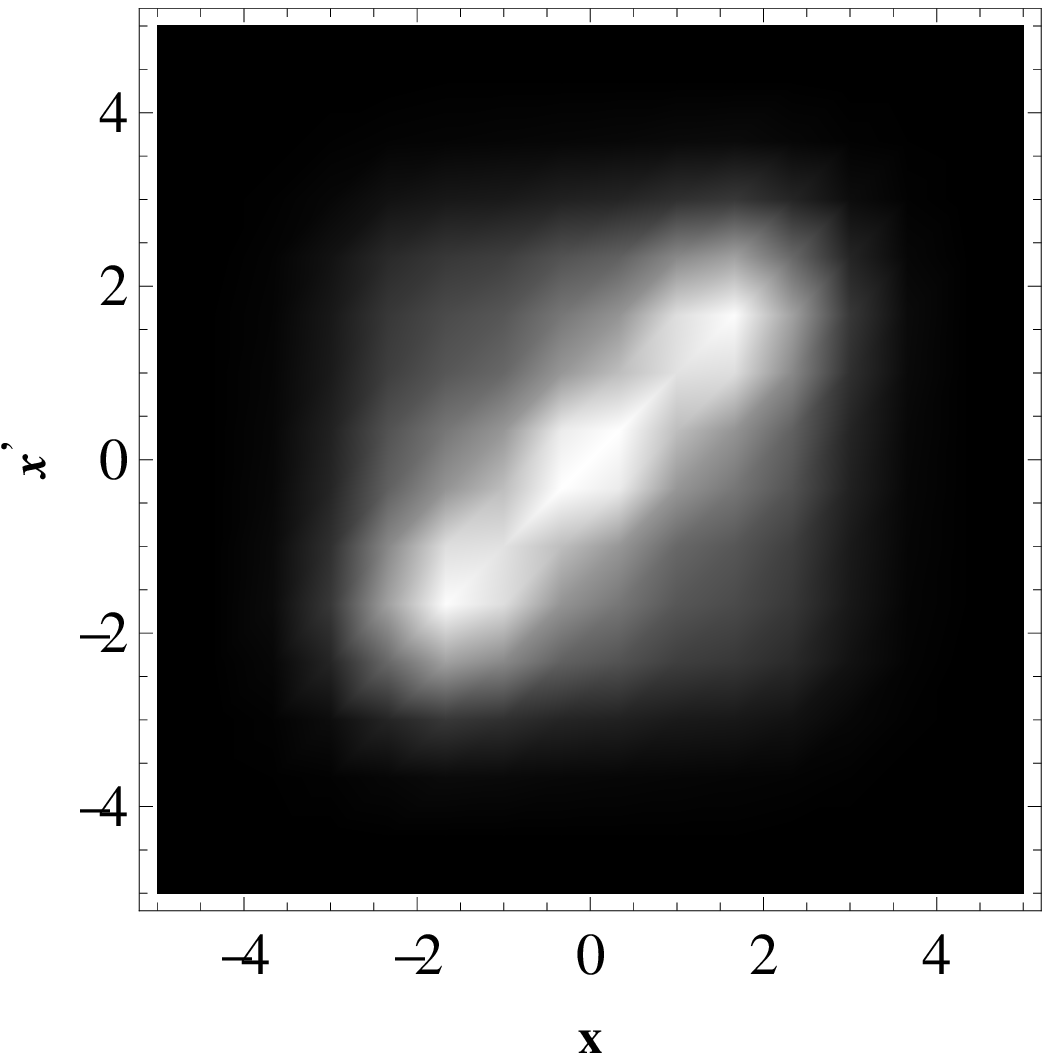}
\includegraphics[width=0.143\textwidth]{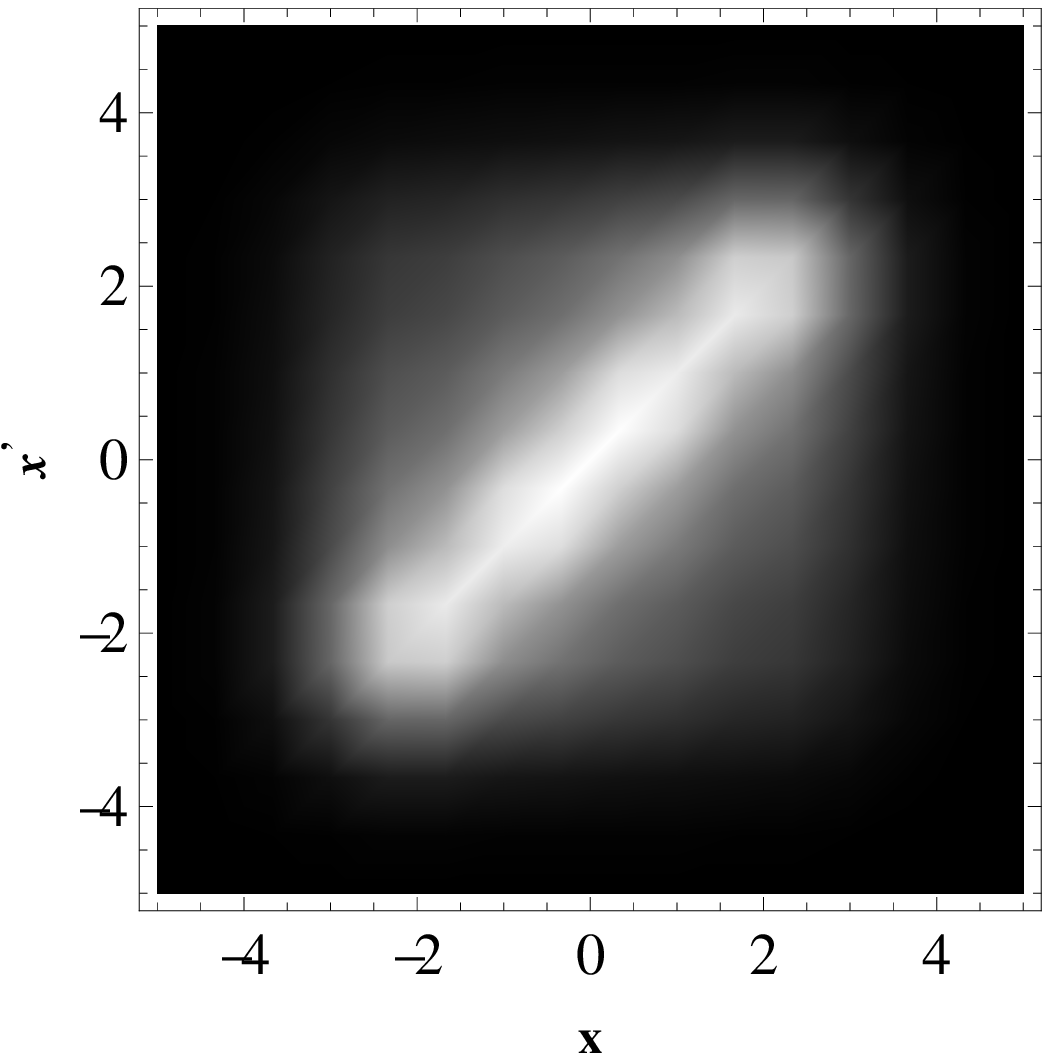}
\includegraphics[width=0.143\textwidth]{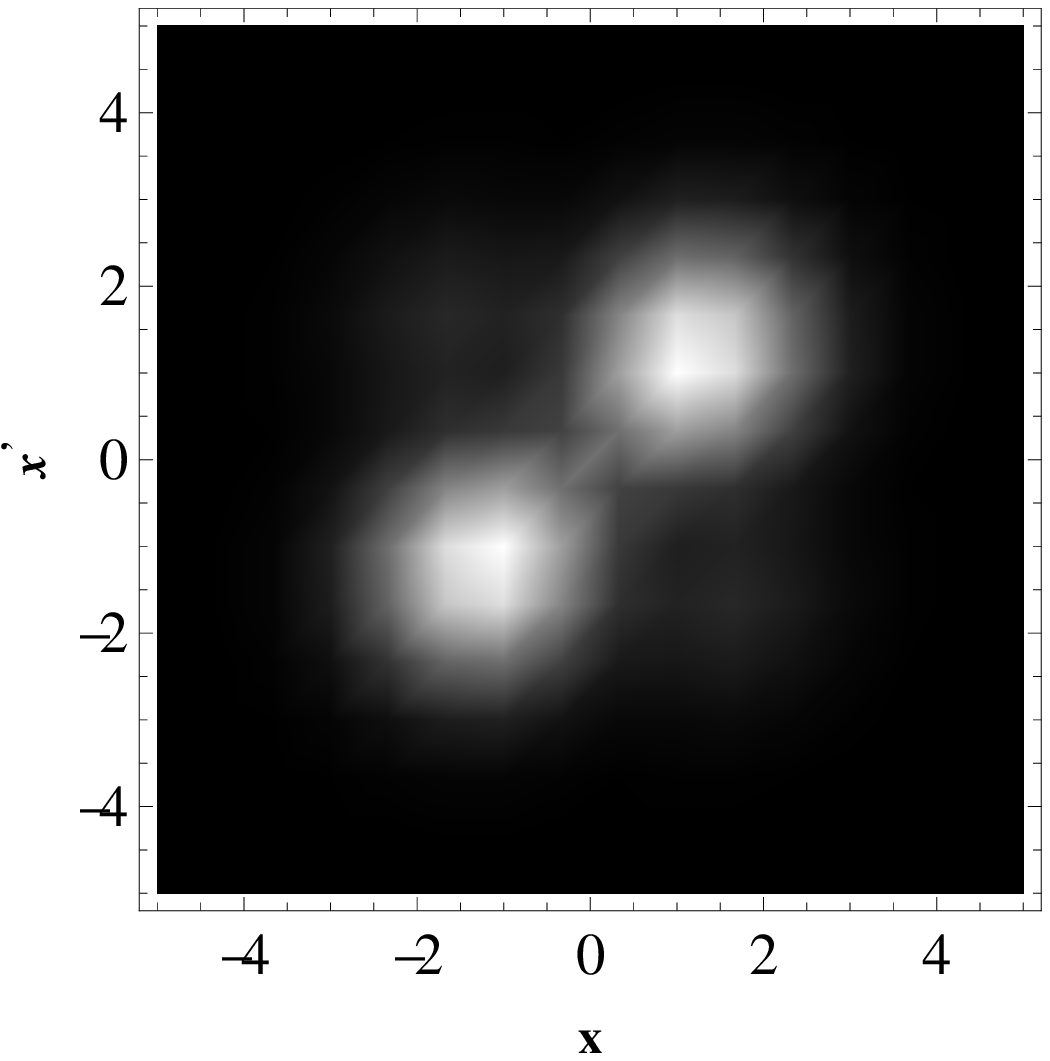}
\includegraphics[width=0.143\textwidth]{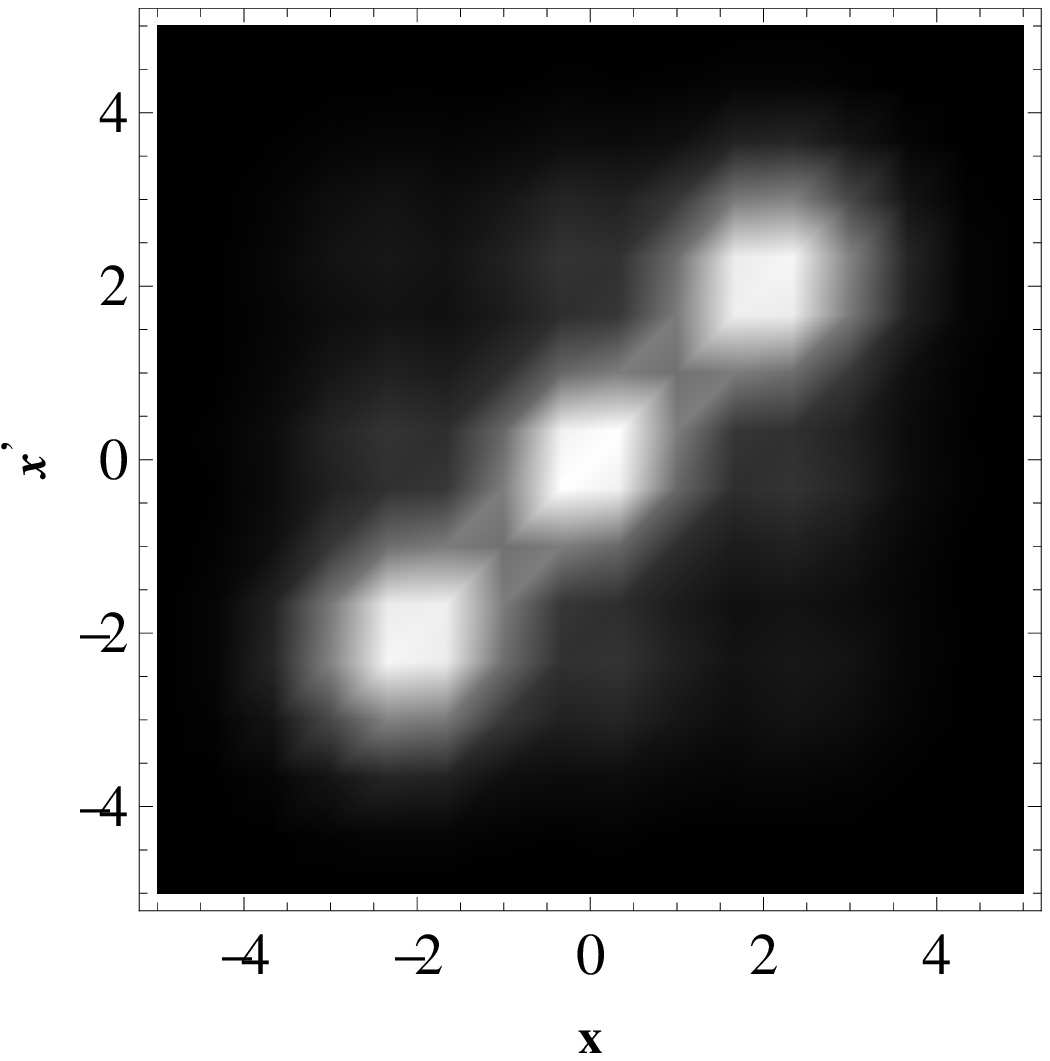}
\includegraphics[width=0.143\textwidth]{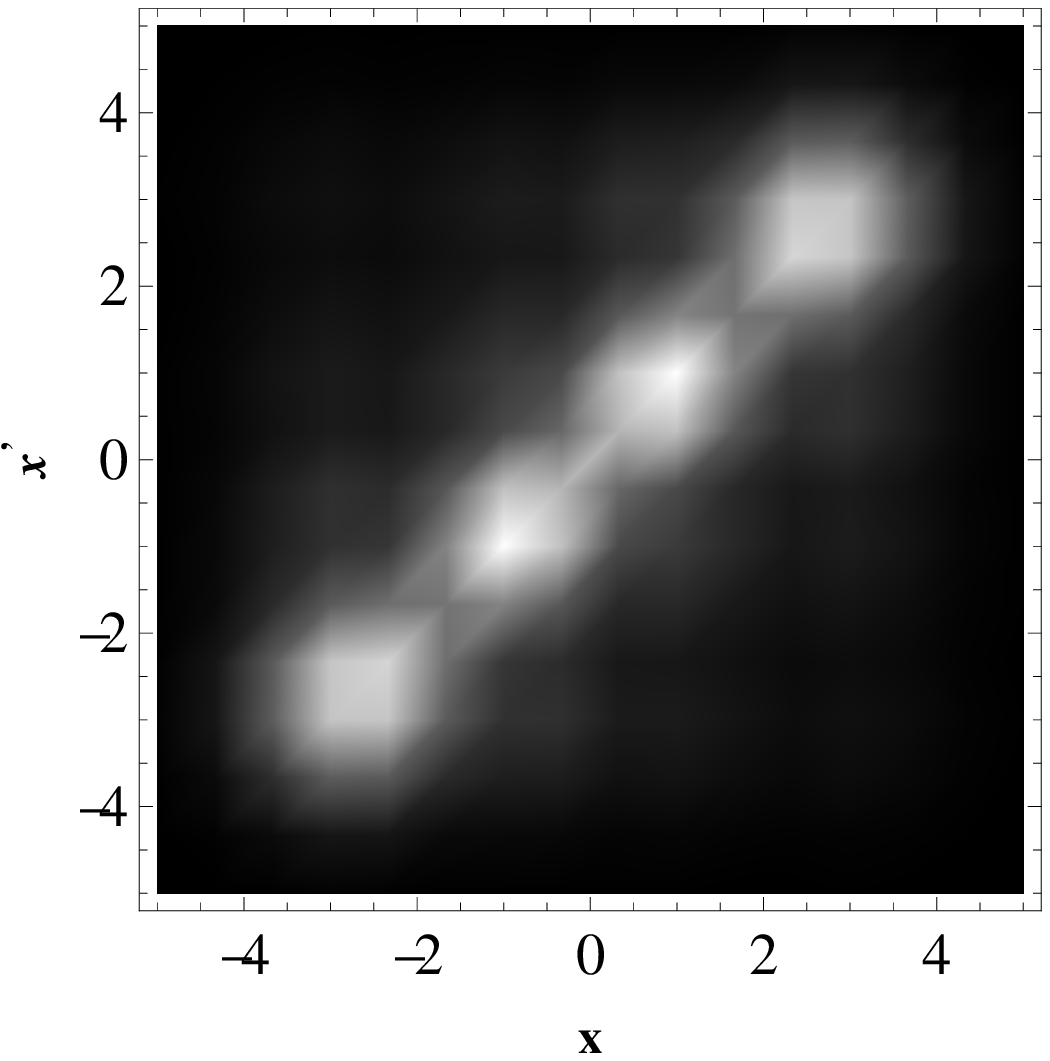}

\end{center}
\caption{\label{fdffklolofghklg:beh} Gray-scale plots of the
effective RDM for systems of $N=2,3$ and $4$ particles with
confinement anisotropy $\epsilon=30$ at various interaction
strengths $g$.}\end{figure}

\begin{figure}[h]
\begin{center}

\begin{picture}(6,6)

\put(72.5,1) {\small $N=4$}
\put(-5.,1) {\small $N=3$}

\put(-80.0,1) {\small  $N=2$}
\put(-125.0,-23) {\scriptsize $g=1$}
\put(-125.0,-100) {\scriptsize $g=2$}
\put(-125.0,-174) {\scriptsize $g=5$}

\end{picture}

\includegraphics[width=0.143\textwidth]{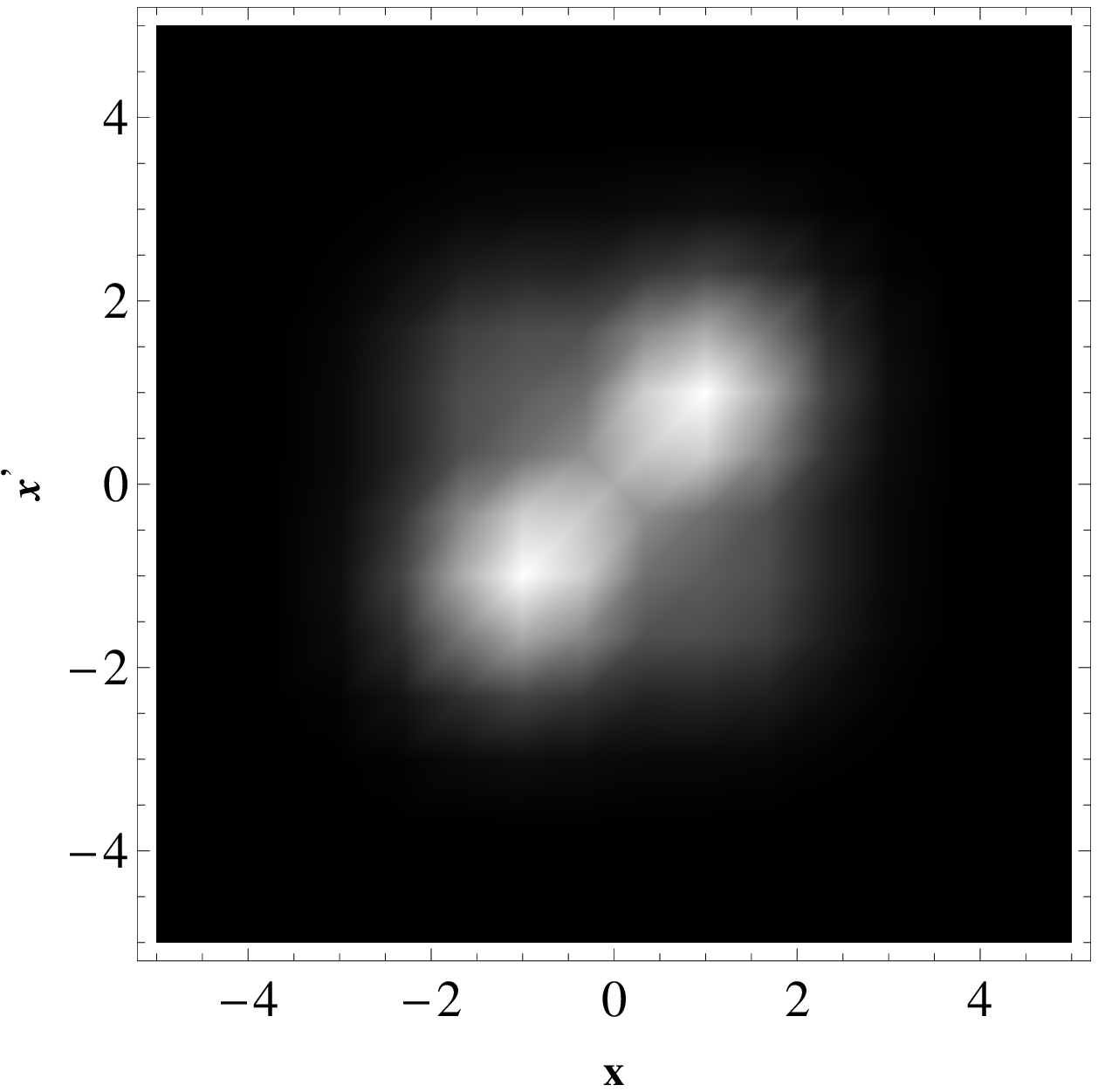}
\includegraphics[width=0.143\textwidth]{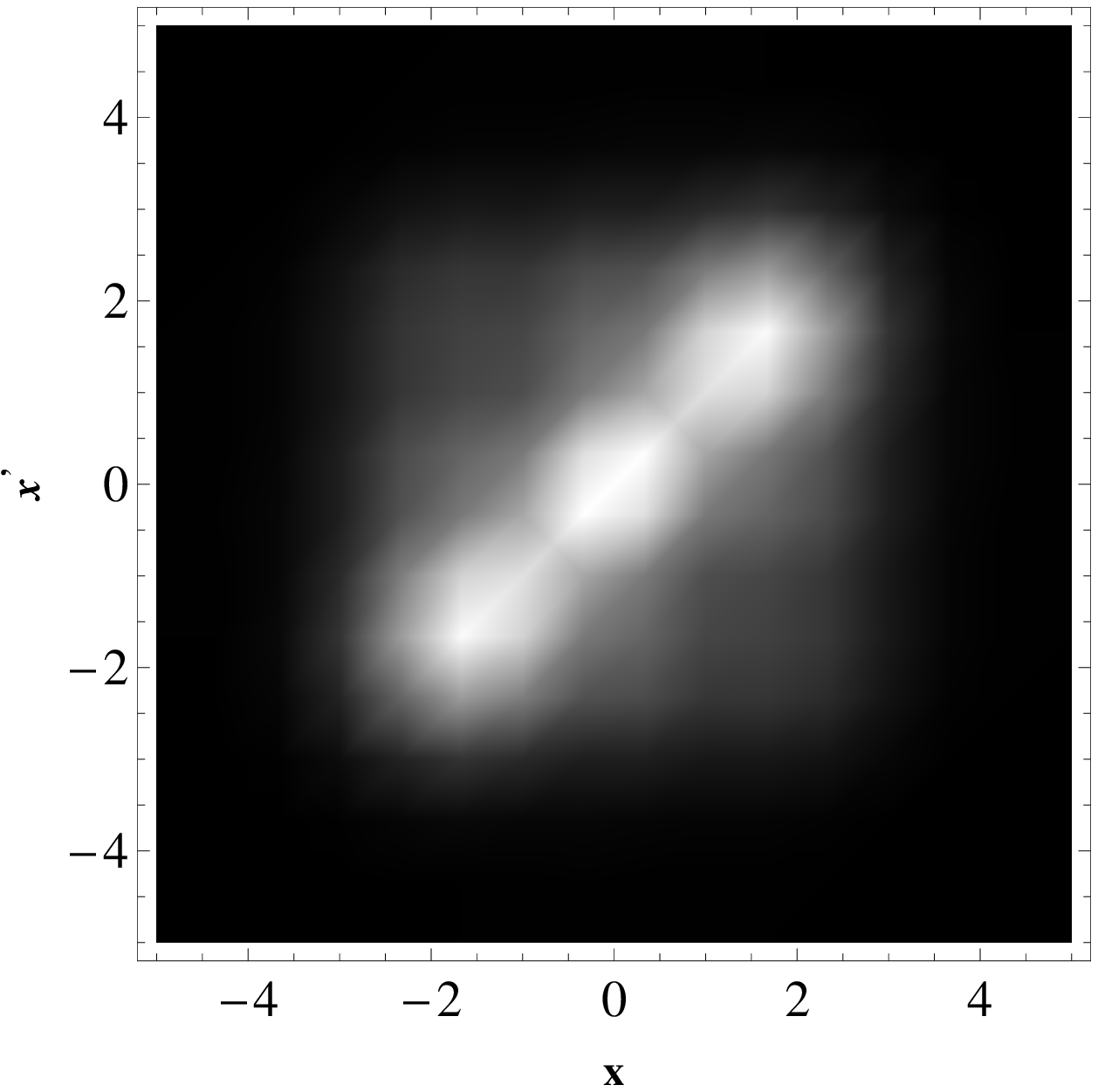}
\includegraphics[width=0.143\textwidth]{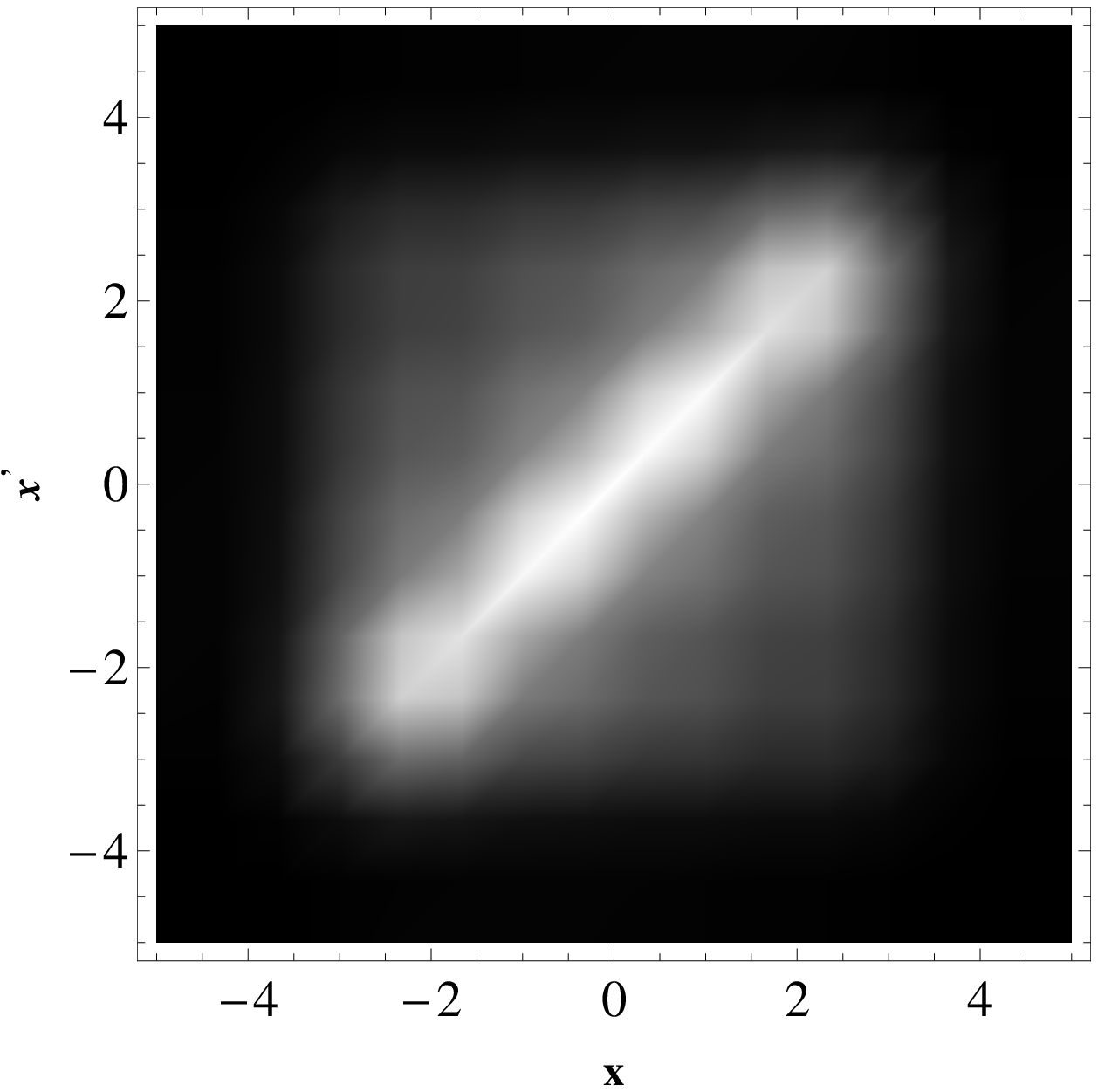}
\includegraphics[width=0.143\textwidth]{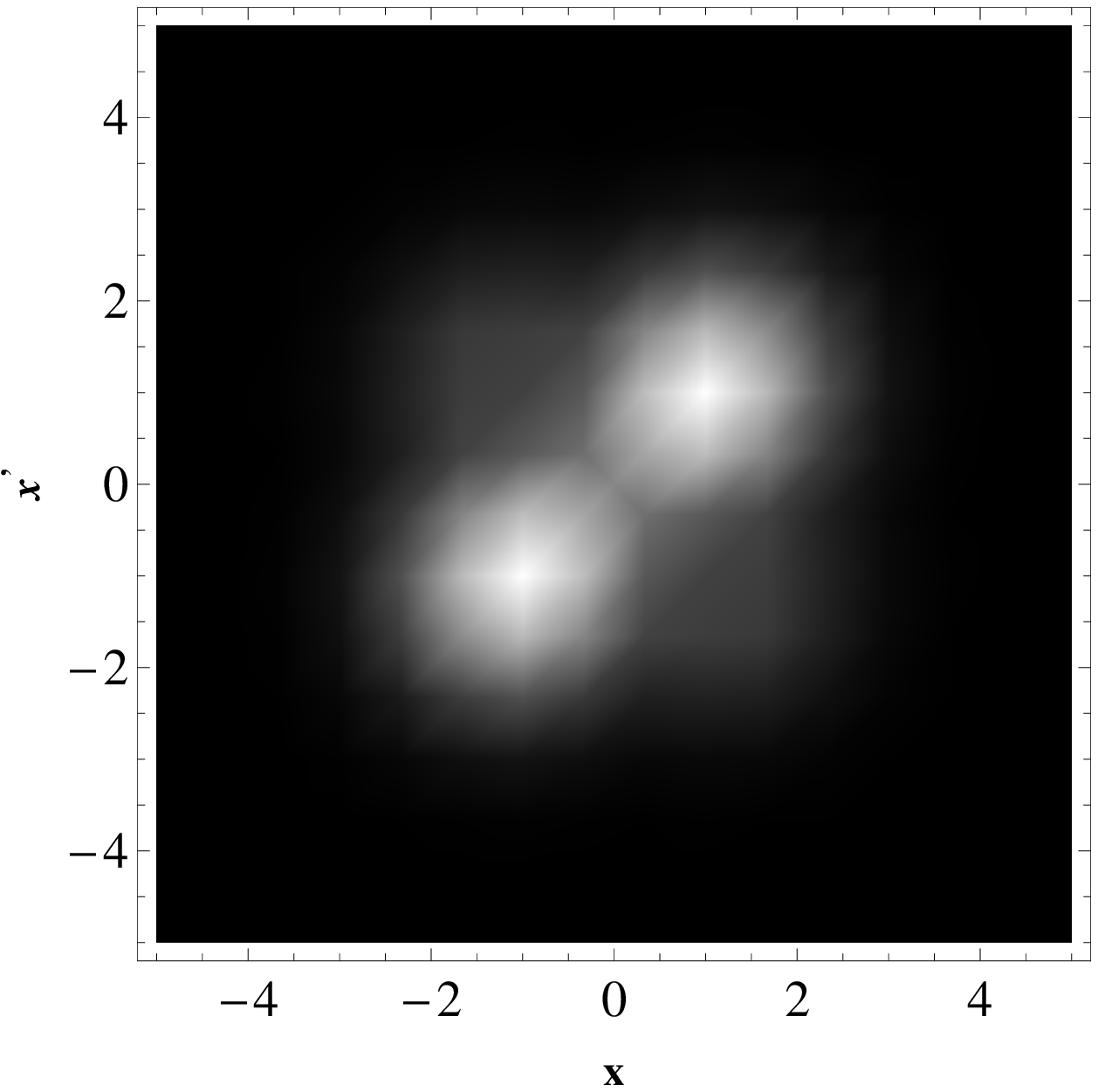}
\includegraphics[width=0.143\textwidth]{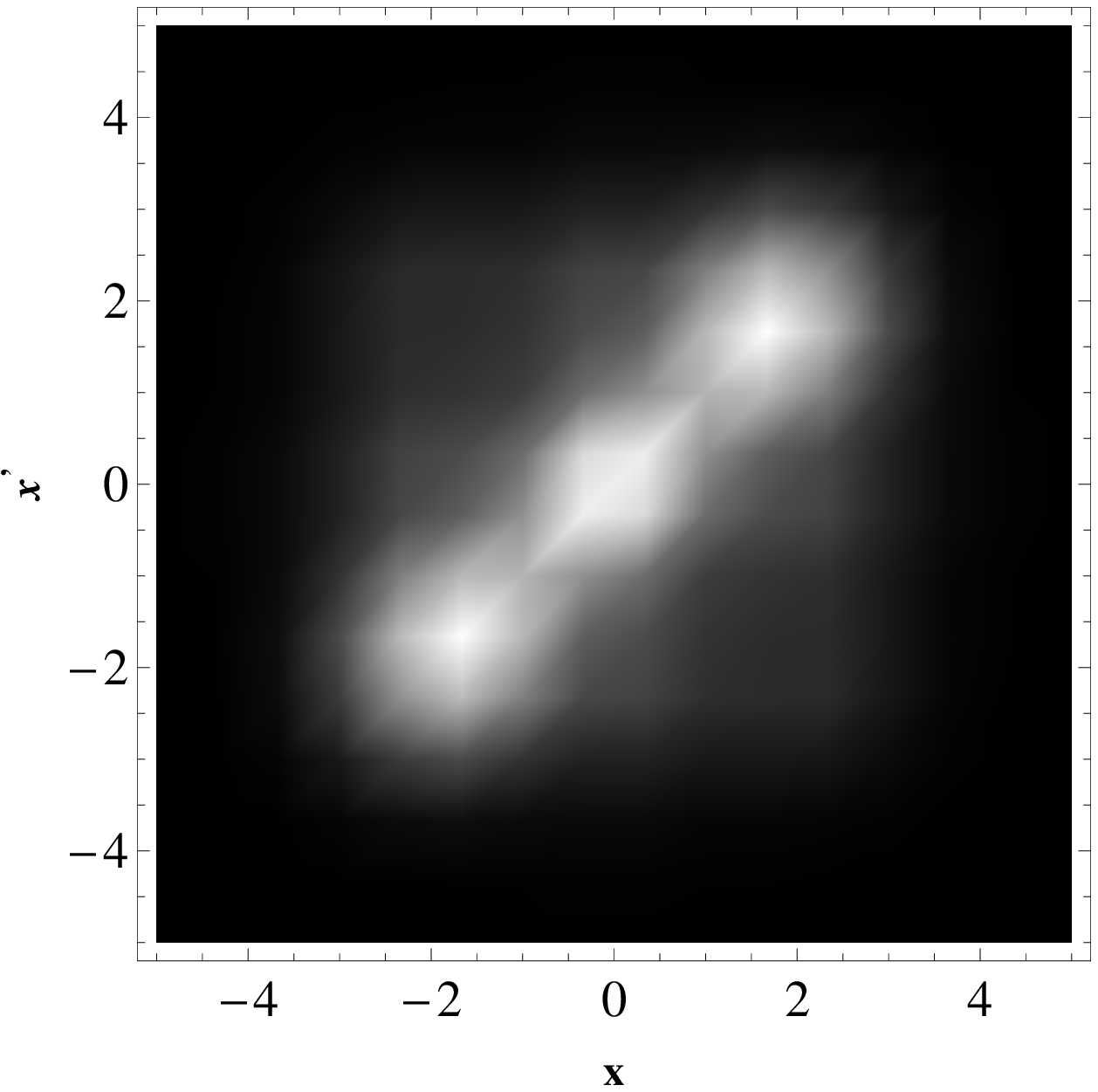}
\includegraphics[width=0.143\textwidth]{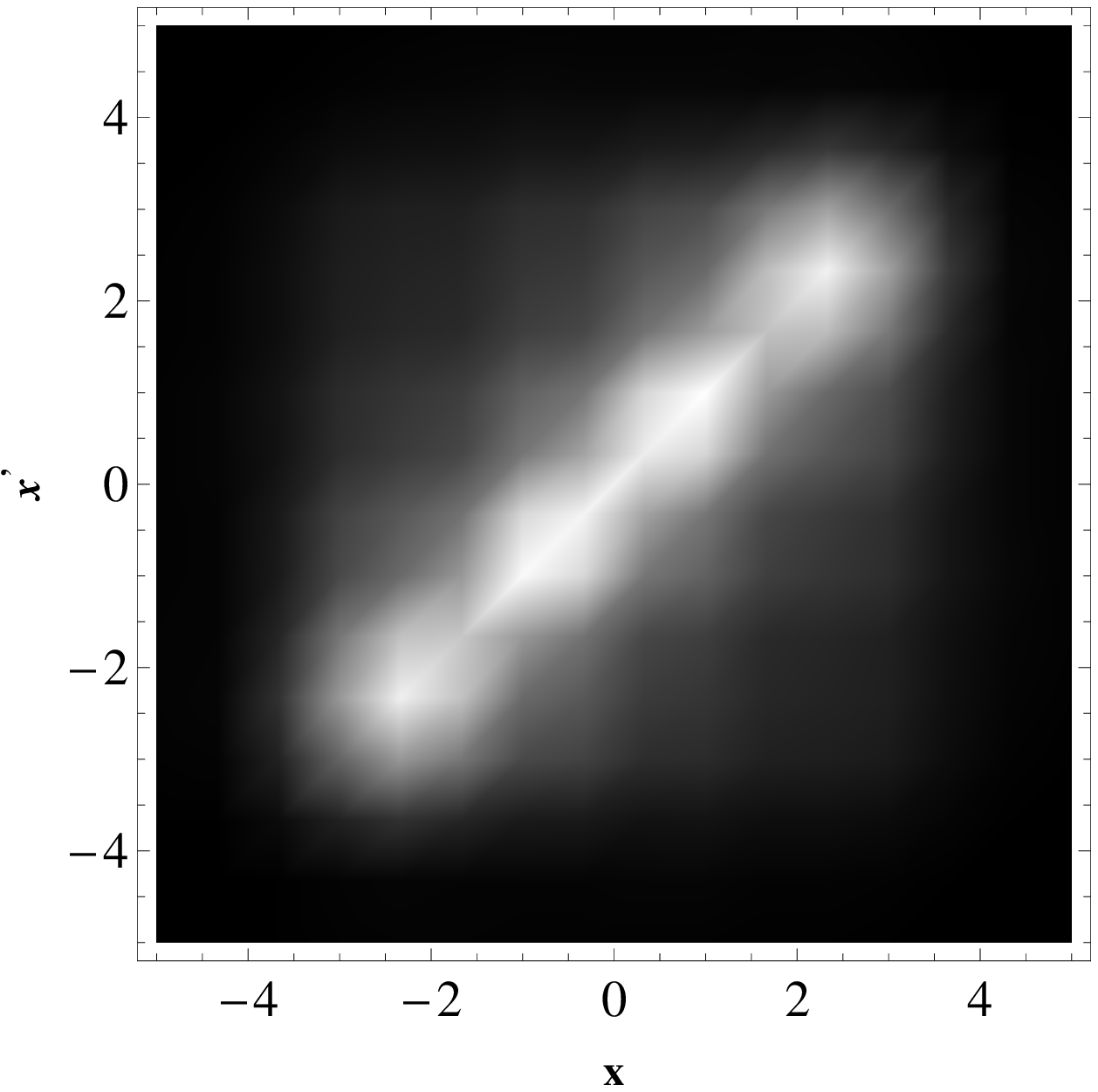}
\includegraphics[width=0.143\textwidth]{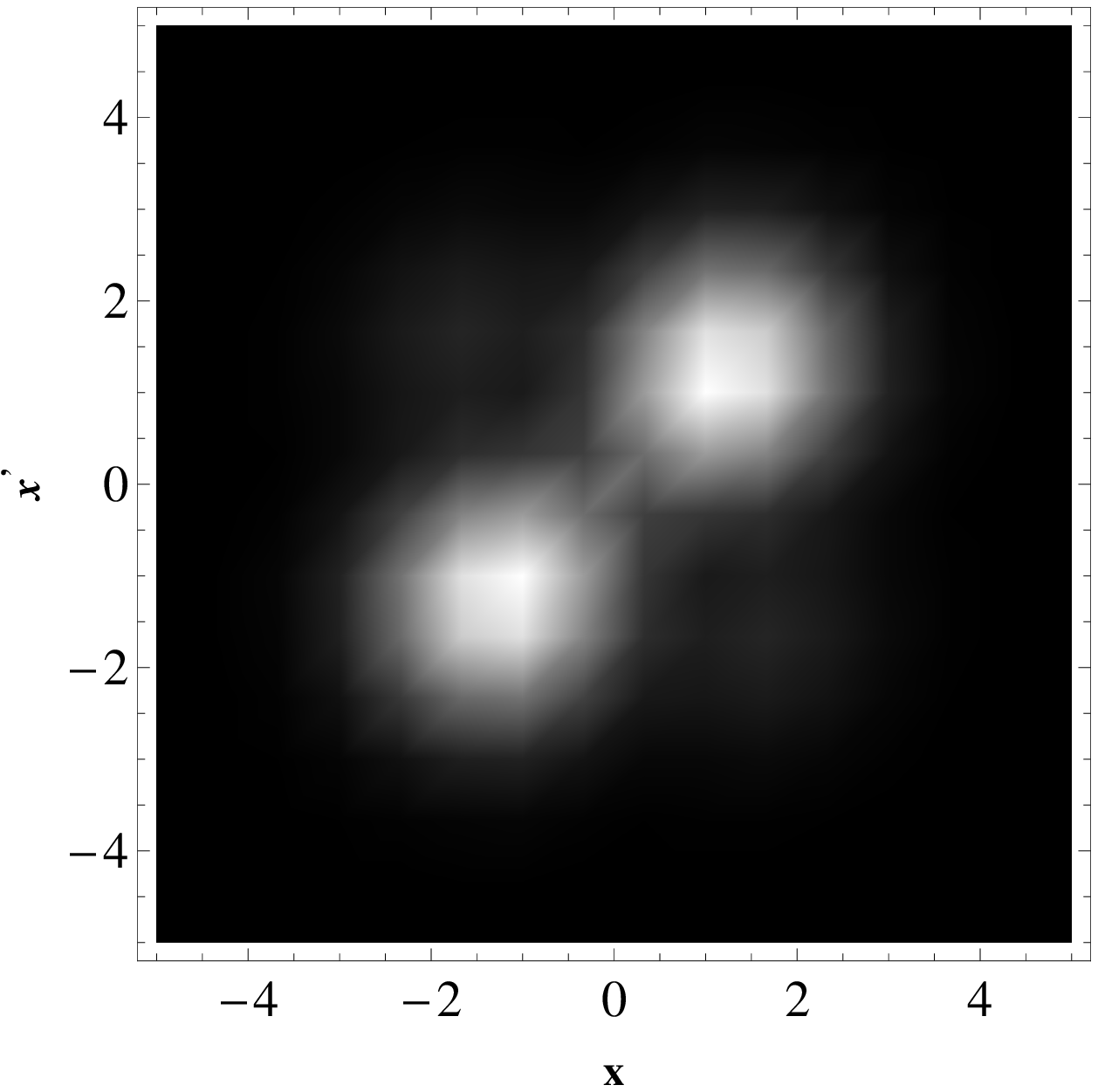}
\includegraphics[width=0.143\textwidth]{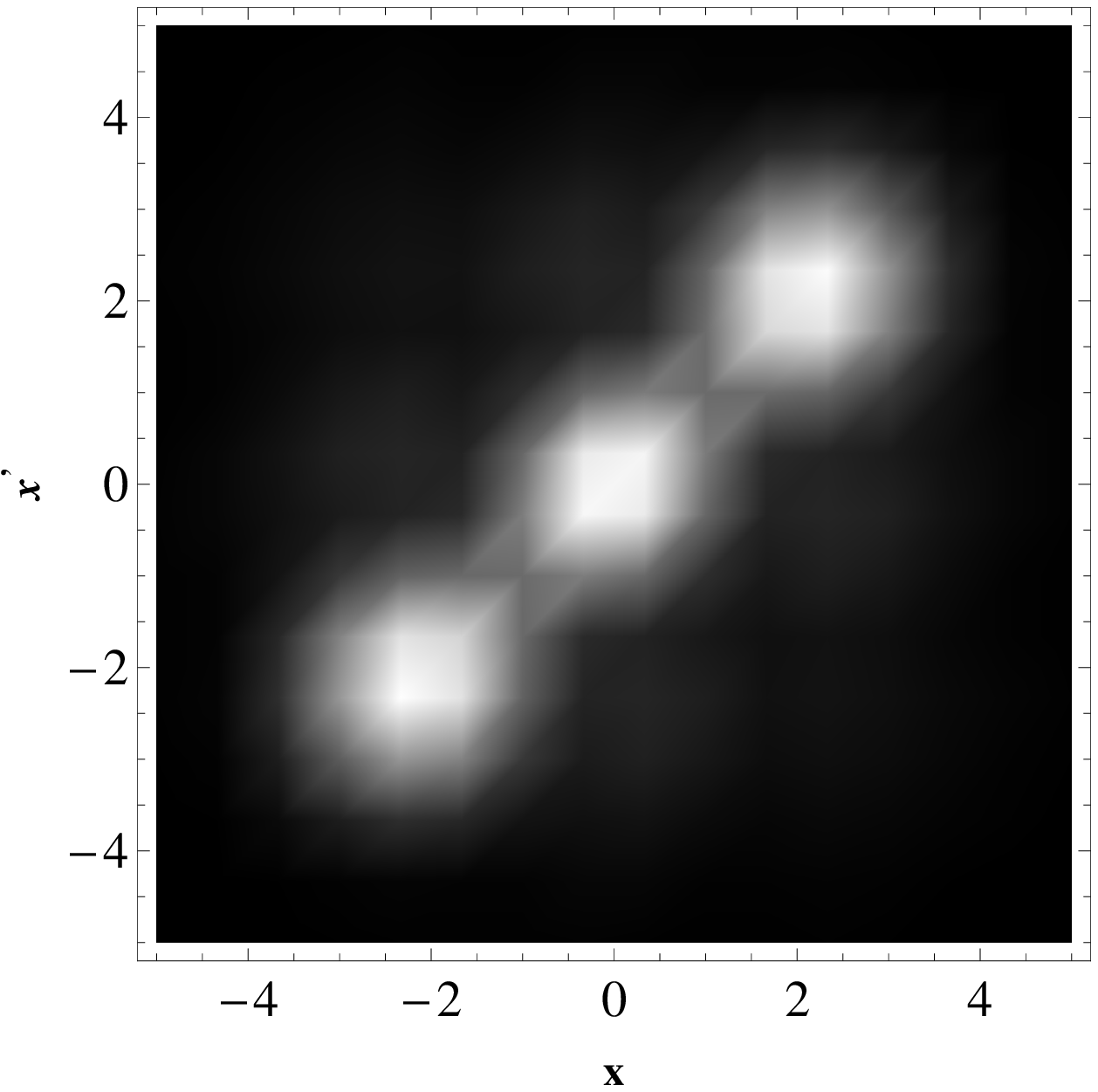}
\includegraphics[width=0.143\textwidth]{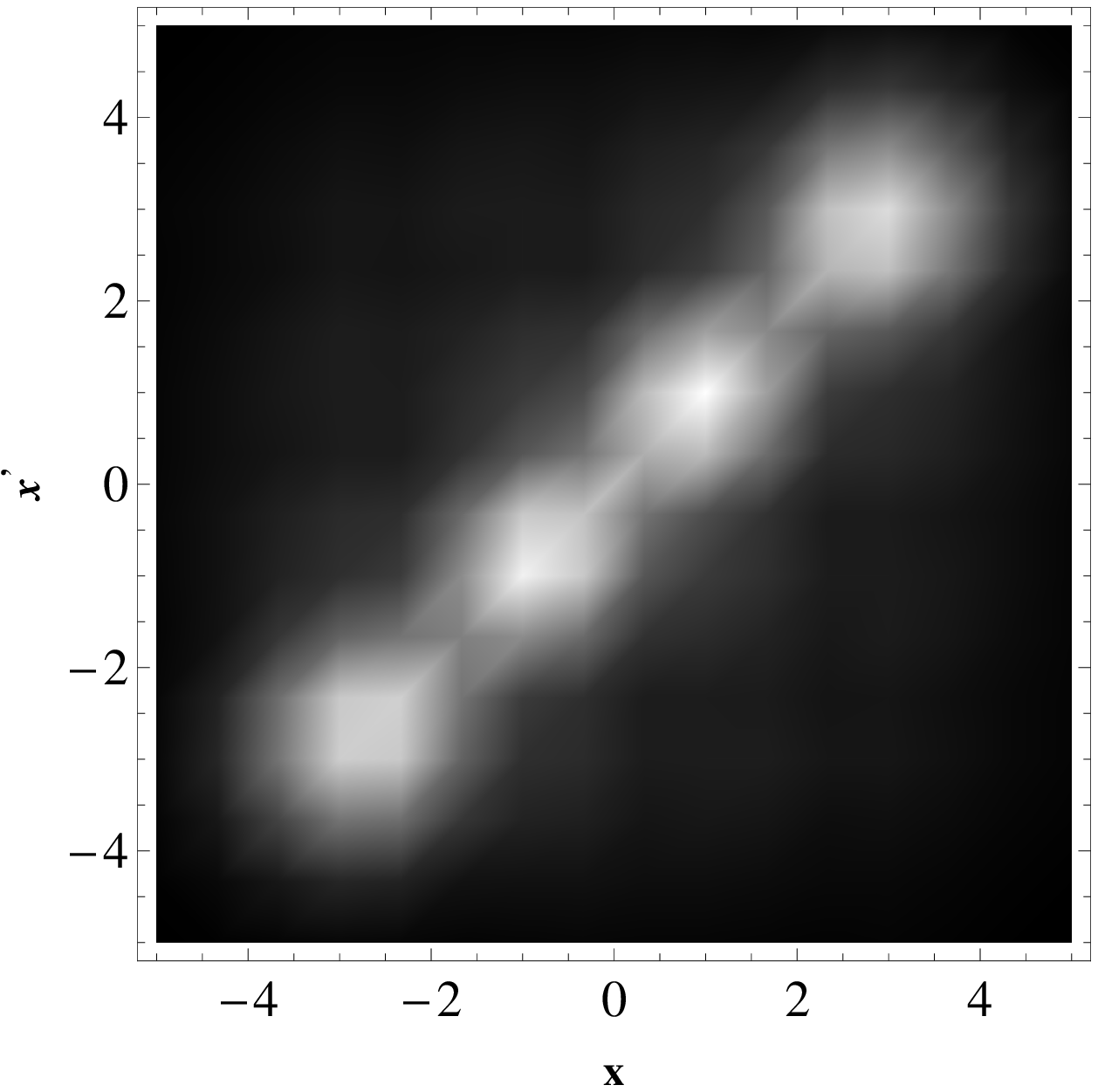}

\end{center}
\caption{\label{infinity} Gray-scale plots of the effective RDM for
systems of $N=2,3$ and $4$ particles with confinement anisotropy
$\epsilon=\infty$ at various interaction strengths $g$.}\end{figure}

Grey-scale plots of $\rho(x,x^{'})$ for the considered systems are shown in Fig.\
\ref{fdffklolofghklg:beh} at three values of the interaction strength
$g$. As one can see, the off-diagonal elements of the RDM diminish
with increasing $g$, which indicates a loss  of spatial coherence in
the system. This is due to the repulsive nature of the interaction
encouraging localization of the particles, which tend to separate
from each other. A clear deviation from the circular structure
$\rho(x,x^{'})={ {\pi}^{-{1\over 2}}}e^{-{(x^2+{x^{'}}^2)\over 2}}$
of the noninteracting case ($g=0$) is clearly observed already at
$g=1$. Interestingly enough, the results of Fig.\
\ref{fdffklolofghklg:beh} show that the onset of crystallization
shows up at $g=5$, independently on the number of particles.

\begin{figure}[h]
\begin{center}

\begin{picture}(6,6)
\put(-102.,-11) {\small $g=1$} \put(17.5,-11) {\small $g=2$}

\put(-43.5,-104) {\small $g=5$}

\end{picture}

\includegraphics[width=0.227\textwidth]{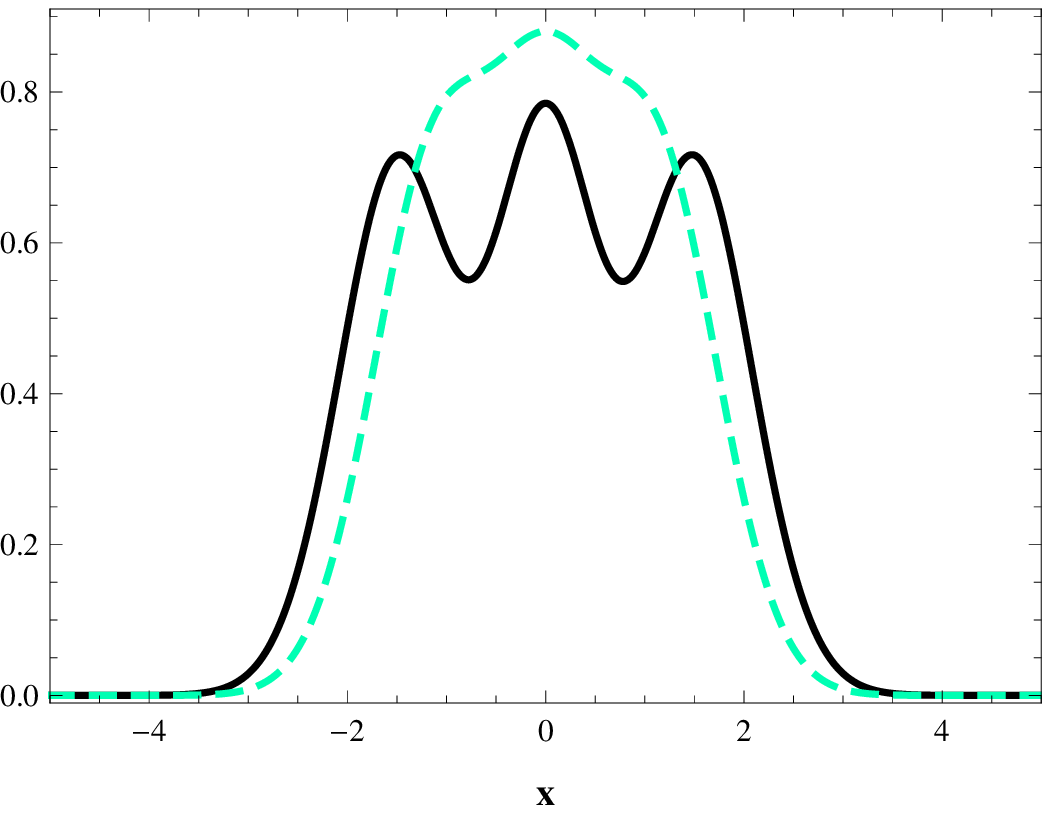}
\includegraphics[width=0.227\textwidth]{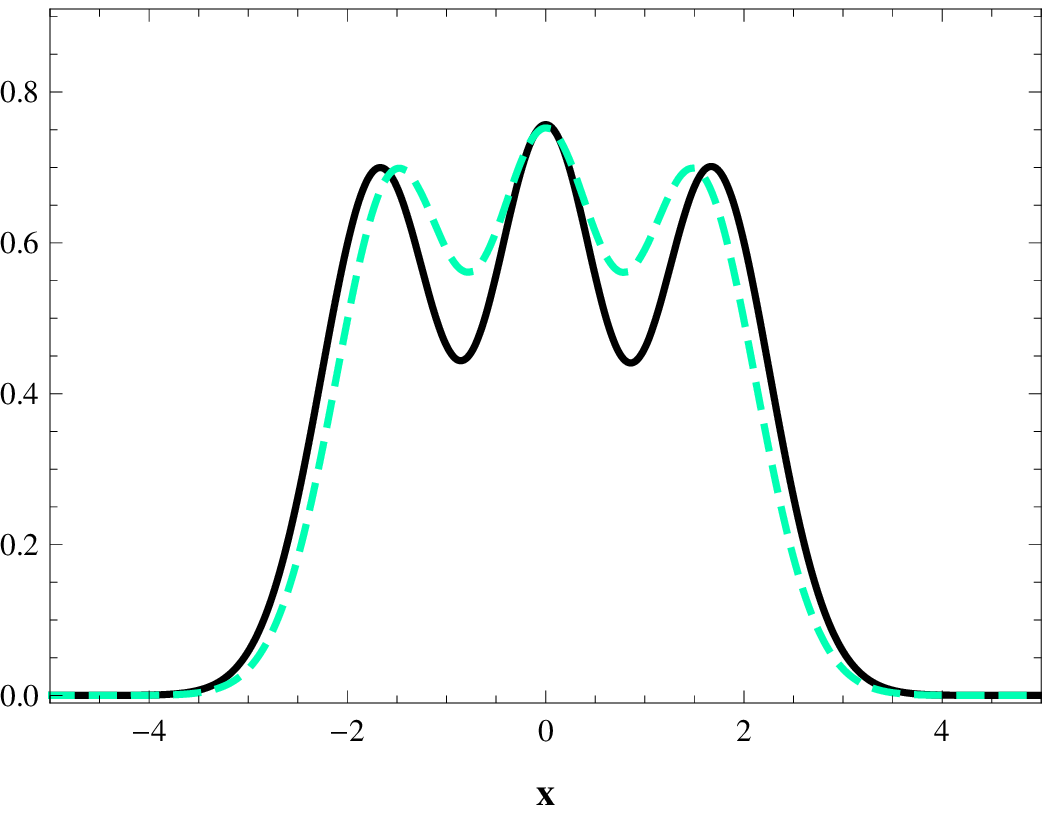}
 \includegraphics[width=0.227\textwidth]{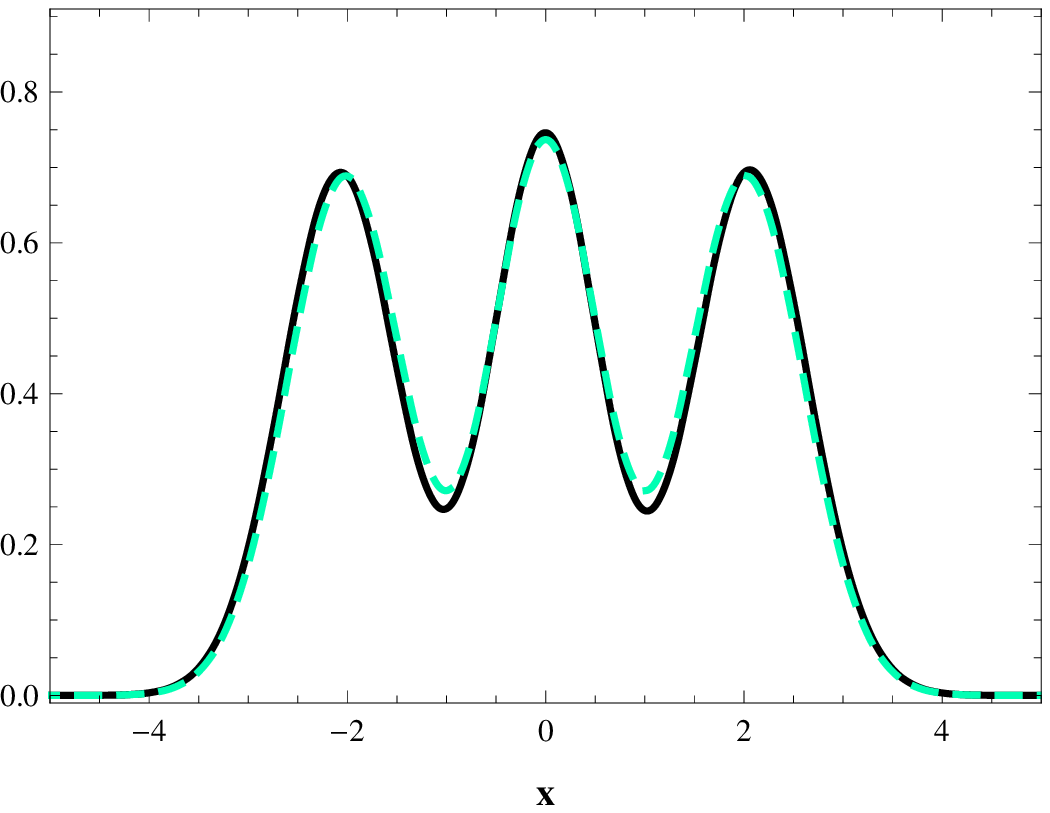}

\end{center}
\caption{\label{fff4444eeklolofghklg:beh}Single-particle density
  of $N=3$ particles confined with anisotropy $\epsilon=30$ (\emph{dashed curve}) and $\epsilon=\infty$ for various interaction strengths $g$.
  }
  \end{figure}

The
results at finite anisotropy $\epsilon=30$ may be compared with those obtained in the strictly 1D limit
that are given in Fig.\ \ref{infinity}. In this case we determined
$\psi_{F}$, and thereby $\psi_{B}$,  using the standard configuration
interaction method based on harmonically trapped single-particle
eigenfunctions $\varphi_{n}^{ho}$. Comparing the results of Fig.
\ref{fdffklolofghklg:beh} with the ones of Fig.\ \ref{infinity}, one
can notice that the anisotropy parameter  influences the
behaviour of the RDM  only in the regime of small values of
$g$. As a matter of fact, already at $g=5$ the RDM calculated at
$\epsilon=30$ reproduces quite well the one calculated in the
strictly 1D limit, regardless of the number of particles.
We observe that below this value an increase in the anisotropy parameter  has the
effect of an increase in the interparticles distances. Otherwise stated,
the onset of the Wigner crystalization in the strictly 1D limit
appears at value smaller than that in the case of $\epsilon=30$.
To make the above more clear,  we present in Fig. {\ref{fff4444eeklolofghklg:beh} the single particle
densities of the $3$-particle system,
 corresponding to the $N=3$ results of Figs.
 \ref{fdffklolofghklg:beh}  and  \ref{infinity}.

Next, we explore the effect of the anisotropy parameter $\epsilon$ on the correlation
properties of systems containing $N=2,3$ and $4$ bosonic particles. In Fig.\
\ref{fffddeeklolofghklg:beh} the ground-state linear entropy $L$ at
different values of $\epsilon$ is compared with the result obtained
in the fermionized strictly 1D limit of $\epsilon\rightarrow\infty$.
The results are presented as a function of the dimensionless parameter $g$.

\begin{figure}[h!]
\begin{center}
\begin{picture}(10,10)
\put(-50,-16) {\small $N=2$} \put(-50,-153) {\small $N=3$}
\put(-50,-287) {\small $N=4$}
\end{picture}

\includegraphics[width=0.35\textwidth]{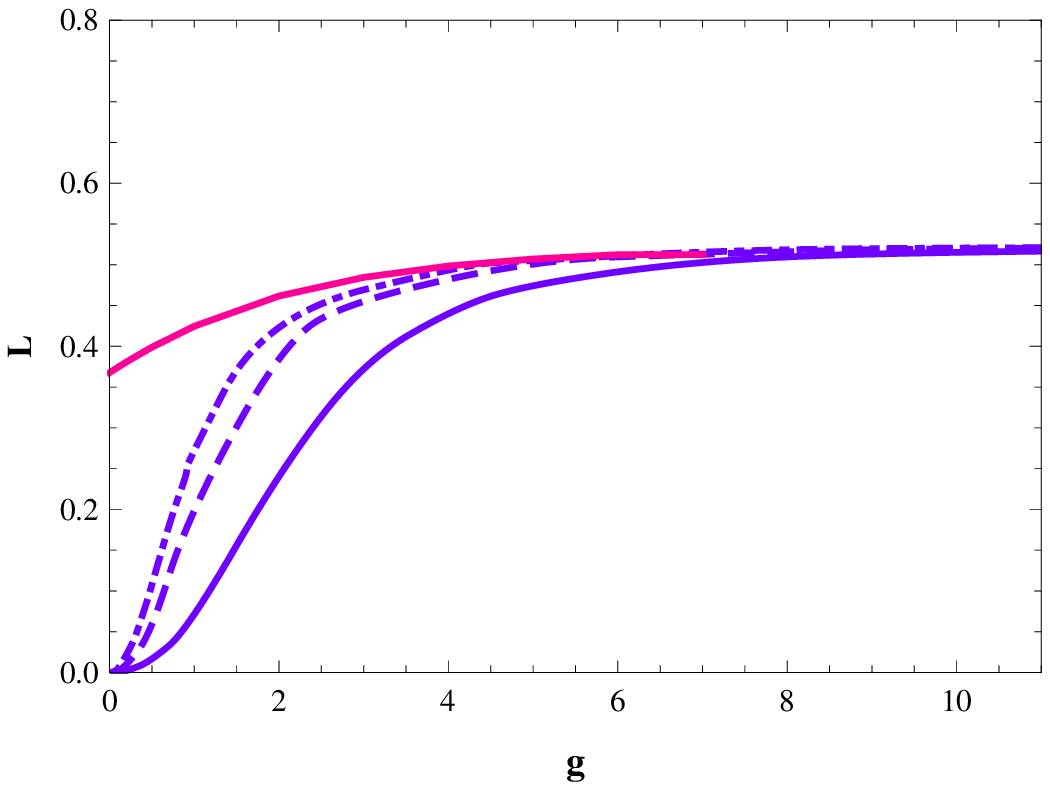}
\includegraphics[width=0.35\textwidth]{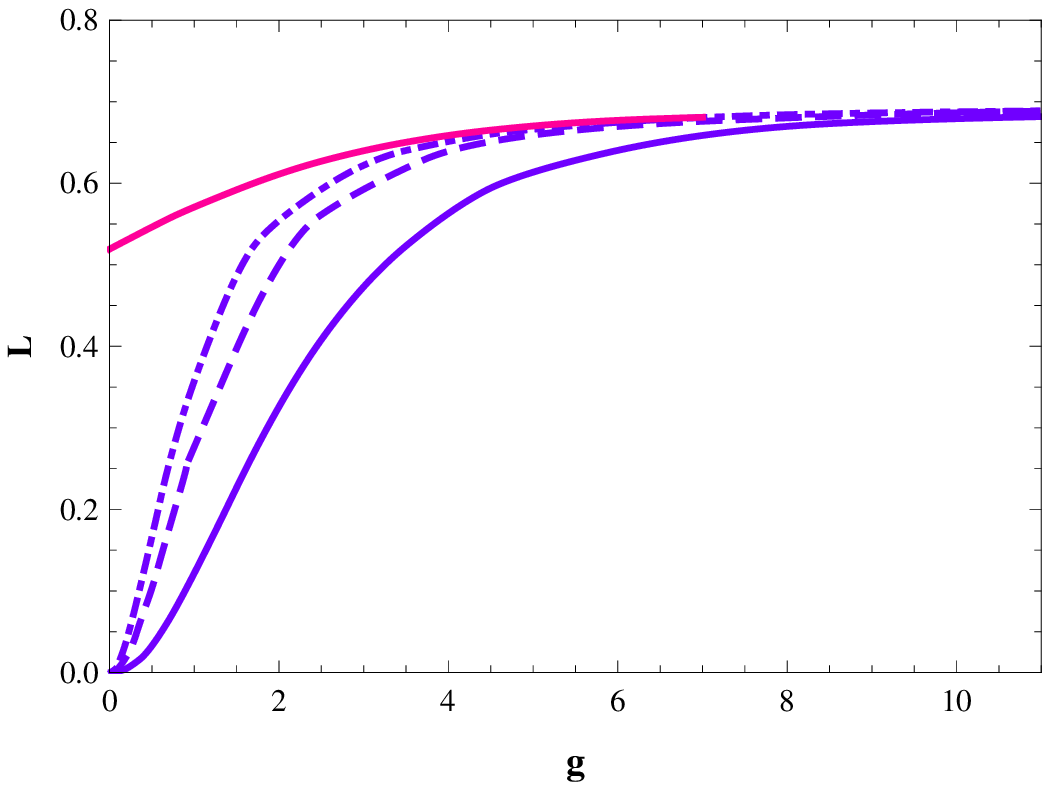}
\includegraphics[width=0.35\textwidth]{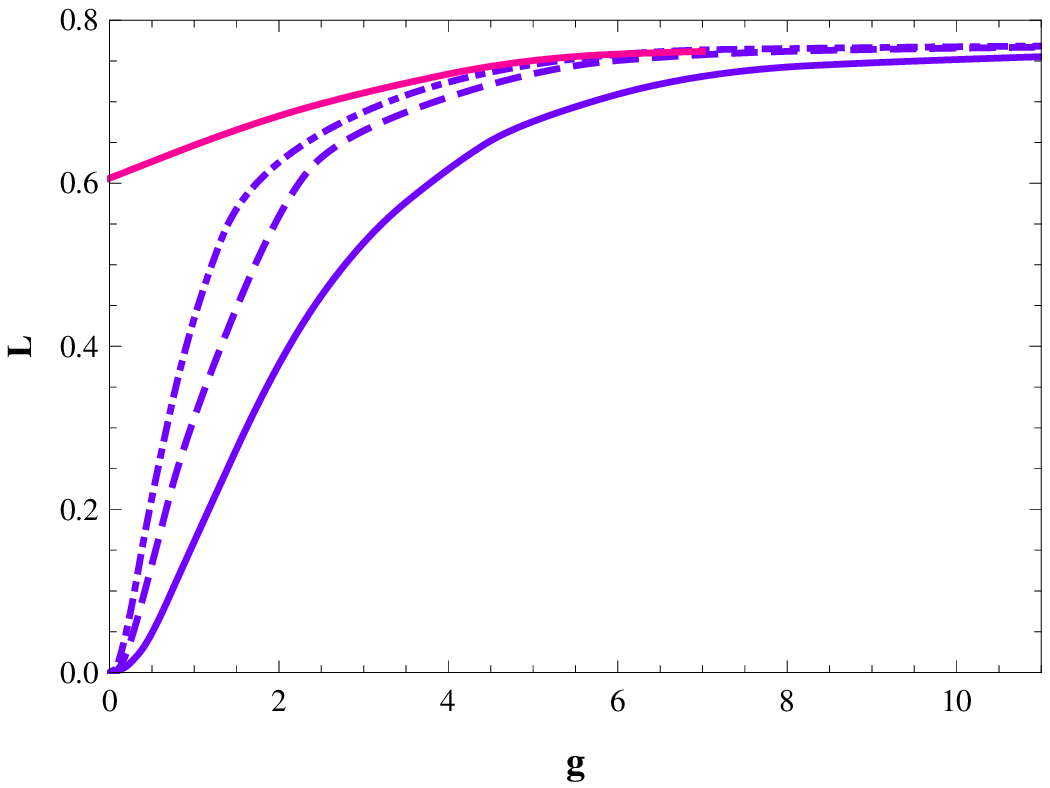}
\end{center}
\caption{\label{fffddeeklolofghklg:beh}Linear entropy in the
ground-state of the system of (\ref{quasio}) for $N=2,3$ and $4$
particles as a function of $g$. \emph{Full curve}, $\epsilon=5$;
\emph{dashed curve}, $\epsilon=30$; \emph{dot-dashed curve},
$\epsilon=100$. The red line is the linear entropy in the strictly 1D limit.}
\end{figure}
We see that in each considered case, the linear entropy increases
with $g$ and above $g\approx 5$ saturates at a constant
value that is insensitive to the anisotropy parameter $\epsilon$.
This may be attributed to the fact that at large $g$ the average
distances between the particles are large enough so that the
effective interaction exhibits pure Coulomb behaviour. The values of $g$
at which saturation takes place seem to
increase slightly with $N$. A look at Fig.\ref{fdffklolofghklg:beh}
allows us to conclude that they coincide roughly with the
critical values of crystallization. As can be inferred from Fig. \ref{fffddeeklolofghklg:beh}, the linear entropy
$L$ increases with the number of particles in the system, with the
increase getting smaller for larger $N$. The effect becomes less
pronounced at weaker interactions and disappears in the limit of
$g\to 0$, when $L\to 0$, regardless of the number of particles.
At small values of $g$, the entropy $L$ depends
strongly on the anisotropy, being larger at larger $\epsilon$.
Again, this can be qualitatively understood by referring  to the
 distances of the particles, namely
  they  are small at small values of $g$
  and are thus in the regime where the effective interaction potential (\ref{eff}) strongly depends on $\epsilon$.

The entropy in the strictly 1D limit of $\epsilon\rightarrow\infty$ has been calculated by taking
into account the Bose-Fermi mapping (\ref{lld}). In the limit of $g\rightarrow 0$, the wavefunction $\psi_{F}$
reduces to a single Slater determinant and (\ref{lld}) takes the
form \be \psi(x_{1},x_{2},...,x_{N})={1\over
\sqrt{N!}}|det_{n=0,j=1}^{N-1,N}(\varphi_{n}^{ho}(x_{j}))|.\label{tgfun}\ee
The values of $L_{1D}$ are calculated to be about $0.36, 0.51$ and
$0.6$ for $N=2,3$ and $4$, respectively. The occurrence of
fermionization for $g\ne 0$ results in the discontinuity of the
linear entropy in the point $g=0$. Importantly, we observe that the smaller is $g$ and/or $N$
the  larger is the anisotropy parameter at which  the linear entropy
of the quasi-1D system reaches the fermionic behaviour of the strictly 1D gas ($\epsilon\rightarrow \infty$).

The asymptotic value of $L_{1D}$ at $g\rightarrow \infty$ can be calculated analytically in the case of $N=2$,
using the harmonic approximation
which becomes exact in this limit \cite{kosPLA}. The calculation
analogous to that performed by us in the 2D case \cite{kosPLA} gives
\be L^{g\rightarrow \infty}_{1D}=1-\sqrt{-{3\over
2}+\sqrt{3}}\approx 0.518\label{pl},\ee which is in agreement  with
our  numerical result. For $N=3$ and $N=4$ the values of
$L_{1D}^{g\rightarrow \infty}$ are found numerically to be about
$0.68$ and $0.77$, respectively. To fully  reveal the correlation
effects in the strong interaction limit, it would be desirable to
obtain the full dependence of $L^{g\rightarrow \infty}$ on $N$.
This is a topic for future investigation.

\section{Summary}\label{4}
We investigated the ground-state correlation properties of the
systems composed of two, three, and four particles in a strongly
anisotropic harmonic trap. Within the one-mode approximation, we
studied the influence of both the number of particles $N$ and the
anisotropy parameter $\epsilon$ on the correlation properties of the
systems in the whole range of repulsive interaction strength $g$. As
a general trend we found that the linear entropy $L$ increases with
increasing $g$. At small $g$ (weak interaction and/or strong confinement), the entropy of the considered
systems depends heavily on the anisotropy of the trap, being larger
at higher anisotropy. Linear entropy is the largest in the limit of
$\epsilon\rightarrow\infty$, when fermionization takes place for any
$g\neq 0$. At large $g$ (strong interaction and/or weak confinement), the
entropy $L$ saturates at a value that does not depend on $\epsilon$
and is greater the larger is $N$. The value of $g$ at which
saturation takes place hardly depends on the number of particles,
shifting only slightly towards larger values with increasing $N$.

The practical realizations of the model discussed in our work can be achieved experimentally in linear ion traps
where the confining forces in the longitidunal direction are much softer than the radial confinement. In the case of singly charged ions
the parameter $g={k e^2}\sqrt{{m\over \omega_{x}\hbar^{3}}}$, where $k$ is the Coulomb constant and $m$ is the ion's mass, can be controlled by the axial trapping frequency $\omega_{x}$. In current experiments the values of $\omega_{x}$ are below $1 MHz$, which corresponds
to the crystalline phase with micrometer distances between the ions~\cite{lin} and a value of $g$ larger than $10^6$. In this regime, the
correlation depends very weakly on anisotropy and the strictly one-dimensional approximation works well even at moderate values of $\epsilon$.

We think a more extensive analysis of the effects of
anisotropy for a larger number of particles is desirable to get a
deeper insight into the properties of strongly correlated
bosons with Coulomb interaction.


\begin{thebibliography}{99}
\bibitem{fab}L. Jacak, P. Hawrylak, and A. W\'ojs, Quantum Dots, Springer, Berlin, 1997.
\bibitem{ion} D. J. Wineland, et al., Phys.
Rev. Lett. 59, 2935 (1987).
\bibitem{blu}T. Schneider and R. Bl\"{u}mel, J. Phys. B: At. Mol. Opt. Phys. 32,
5017 (1999).
\bibitem{bao}Y. He and C. Bao, J. Phys. B: At. Mol. Opt. Phys. 34,
1641 (2001).
\bibitem{nboson}Y. Kim and A. Zubarev, Phys. Rev. A 64, 013603 (2001).
\bibitem{gonz}  A. Gonzalez, B. Partoens, A. Matulis, and F. Peeters, Phys.
Rev. B 59, 1653 (1999).
\bibitem{int6} M. Olshanii, Phys. Rev. Lett. 81, 938 (1998).
\bibitem{tg0}M. D. Girardeau, E. M. Wright, and J. M. Triscari, Phys. Rev. A 63,
033601 (2001).
\bibitem{1}X. Yin et al., Phys. Rev. A 78, 013604 (2008).
\bibitem{2}D. Murphy, J. McCann, J. Goold, and T. Busch, Phys. Rev. A76, 053616 (2007).
\bibitem{3} T. Sowi\'{n}ski et al., Phys. Rev. A 82, 053631 (2010).
\bibitem{4}C. Matthies, S. Z\"{o}llner, H.-D. Meyer, and P. Schmelcher, Phys. Rev.
A 76, 023602 (2007).
\bibitem{1d} B. Sun, D. Zhou, and L. You, Phys. Rev. A 73, 012336 (2006).
\bibitem{redu}A. Coleman and V. Yukalov, Reduced Density Matrices, (Springer, Berlin, 2000).
\bibitem{lin3} R. Ya\~{n}ez, A. Plastino, and J. Dehesa, Eur. Phys. J. D 56 (2010) 141.
\bibitem{helium}J. S. Dehesa et al., J. Phys. B: At. Mol. Opt. Phys. 45
(2012) 015504.
\bibitem{manz} D. Manzano, et al.,  J. Phys. A: Math. Theor. 43  (2010)
275301.
\bibitem{kosPLA} P. Ko\'scik, A. Okopi\'nska,
Phys. Lett. A 374, 3841 (2010).
\bibitem{astr}G. E. Astrakharchik and M. D. Girardeau, Phys. Rev. B 83, (2011)  153303.
\bibitem{difu}R. Kosloff and H. Tal-Ezer, Chem. Phys. Lett. 127, 223 (1986).
\bibitem{lin} M.G.Raizen et al., Phys. Rev. A 45, 6493 (1992).
\end{thebibliography}
\end{document}